\documentclass[10pt,preprint]{aastex}
\usepackage{epsf,epsfig}

\def\aap{A\&A}
\def\apjl{ApJ}

\newcommand{\Om}{\mbox{$\Omega_M$}}

\begin{document}
\pagestyle{plain}
\setlength{\baselineskip}{0.95\baselineskip}

\title{Probing the Early Universe with the S-Z Effect}

\author{Marshall~Joy\footnote{Dept. Space Science, SD50,
NASA/Marshall Space Flight Center, Huntsville, AL 35812.  e-mail:
marshall.joy@msfc.nasa.gov} ~and
John~E.~Carlstrom\footnote{Department of Astronomy and
Astrophysics, University of Chicago, Chicago, IL 60637.  e-mail:
jc@hyde.uchicago.edu}}

The cosmic microwave background radiation (CMBR) we observe
today provides a window to an early stage in our universe's
evolution, when the expanding universe had cooled to the point
that free electrons and ionized nuclei recombined to form atoms.
Before recombination, scattering between photons and free
electrons was frequent, and the distance that light could
penetrate was small; afterward, with free electrons out of
circulation, the universe became largely transparent to light.
Small variations in the CMBR intensity trace small perturbations
in the primordial matter density, which have been amplified by
gravitational forces to form the magnificent, complex structures
that make up the present-day universe.

In certain massive objects, however, interactions between CMBR
photons and free electrons continue to play an important
cosmological role. The largest gravitationally collapsed
structures in the universe are clusters of galaxies with masses
up to 100,000 times greater than the mass of our galaxy, the
Milky Way. At optical wavelengths, clusters are beautiful
objects consisting of thousands of galaxies, each containing
billions of stars, all bound together by a strong gravitational
field. The galaxies and stars, however, only account for a few
percent of the total mass. Most of the normal (baryonic) matter
resides in the hot ($\sim100$ million K) gas that permeates the
galaxy cluster. When CMBR photons interact with the free
electrons in this ionized gas, a unique feature--the
Sunyaev-Zel'dovich (SZ) effect--is imprinted on the spectrum of
the microwave background. This feature proves to be of
fundamental importance for cosmology.

The interaction of a CMBR photon with a hot cluster electron
will, on average, cause the photon to gain a small amount of
energy. A cluster of galaxies contains a tremendous amount of
gas ($\sim10^{14}$ times the mass of our sun), but the
probability that a CMBR photon will interact with an electron in
the cluster gas is nevertheless small. The SZ effect is
therefore subtle, changing the brightness of the CMBR spectrum
by at most 0.1\%. This spectral distortion has a distinct
signature: In the low-frequency part of the CMBR spectrum, the
SZ scattering process causes the brightness of the CMBR to be
diminished toward galaxy clusters, producing "holes" in the
background radiation field (see the left panel in the first
figure). The scattered photons are shifted to higher energies,
producing an excess in the high-frequency part of the CMBR
spectrum (see the right panel in the first figure).

The SZ effect was first proposed 30 years ago (1, 2) but proved
exceptionally difficult to detect. Accurate measurements are now
possible with experimental techniques developed over the last
decade (3, 4). The SZ effect has been used to independently
determine the expansion rate of the universe (Hubble's constant)
and the matter density of the universe (\Om).

The ability to determine these important cosmological quantities
rests on the fact that the magnitude of the SZ effect is
proportional to the total number of free electrons contained in
the cluster, weighted by their temperature. An accurate measure
of the SZ effect thus leads to an estimate of the cluster gas
mass, provided that the gas temperature is known. This
temperature is obtained from the x-ray emission spectrum of the
hot gas, from which we can infer the kinetic energy and the
total mass required to bind the cluster together. It is then
possible to determine the fraction of normal matter to total
matter contained within galaxy clusters; this fraction is
important because the composition of objects as large and
massive as galaxy clusters should reflect the composition of the
universe as a whole. Finally, the total matter density of the
universe is obtained by scaling the measured baryon density of
the universe (5) by the baryon fraction derived from SZ effect
measurements.

Like other recent techniques, the SZ effect observations
indicate that the mass density of the universe, including the
mysterious dark matter, is quite low: \Om $\sim0.25$ (6, 7). This
measured mass density accounts for only 25\% of the critical
density in a flat universe, which is inferred from recent CMBR
anisotropy measurements (8-10). This suggests that about 75\% of
the present energy density in the universe is in some as yet
undiscovered form.

The expansion rate of the universe, Hubble's constant, can be
determined by combining the SZ effect and x-ray measurements.
The strength of the x-ray emission is proportional to the square
of the gas density, in contrast to the SZ effect, which is
linearly proportional to the gas density. A combination of the
two measurements allows the gas density and cluster distance to
be determined; the expansion rate is then obtained by dividing
the cluster's recessional velocity by its distance. SZ effect
and x-ray observations of a large sample of galaxy clusters
currently under way (3, 4, 11, 12) will provide an independent
measurement of the Hubble constant.

Unlike most emission mechanisms, the brightness of the SZ effect
depends only on the properties of the cluster gas and not on
cluster distance. It will soon be possible to exploit this
powerful and unusual property to explore the distant universe.
The SZ effect will be used to determine the abundance and
evolution of massive galaxy clusters from the time of their
formation to the present, which reflect the underlying
cosmological parameters of the universe (see the second figure).
A large-area interferometric SZ effect survey will be able to
detect massive galaxy clusters at whatever epoch they have
formed (13, 14). Present theories, which assume that the initial
spectrum of density fluctuations has a normal (Gaussian)
distribution, predict that massive clusters should not have
formed at redshifts $z \gtrsim 3$, but this has not yet been
confirmed experimentally (a higher redshift corresponds to a
larger cluster distance and to an earlier period in the
evolution of the universe). If large non-Gaussian fluctuations
existed in the early universe, then clusters will have formed at
earlier epochs than currently predicted. Because the SZ effect
is independent of distance, results of SZ effect surveys will
provide an incisive test of theories of the structure and
evolution of the universe, as well as an independent
determination of fundamental cosmological parameters.\\
\small \\
1. R. Sunyaev, Y. Zel'dovich, Comments Astrophys. Space Phys.
2, 66 (1970).\\
2.  R. Sunyaev, Y. Zel'dovich,, Comments Astrophys. Space Phys.
4, 173 (1972).\\
3. M. Birkinshaw, Phys. Rep. 310, 97 (1999).\\
4. J. E. Carlstrom et al., in Constructing the Universe with
Clusters of Galaxies, F. Durret, G. Gerbal, Eds. (IAP, Paris,
2000),
http://www.iap.fr/Conferences/Colloque/coll2000/contributions/ \\
5. S. Burles, K. Nollett, J. Truran, M. Turner, Phys. Rev. Lett.
82, 4176 (1999).\\
6. S. T. Myers et al., Astrophys. J. 485, 1 (1997).\\ 
7. L. Grego et al., Astrophys. J., in press (see
xxx.lanl.gov/abs/astro-ph/0012067). \\
8. A. D. Miller et al., Astrophys. J. 524, L1 (1999). \\
9. P. de Bernardis et al., Nature 404, 955 (2000). \\
10. S. Hanany et al., Astrophys. J. 545, L5 (2000). \\
11. E. D. Reese et al., Astrophys. J. 533, 38 (2000). \\
12. B. D. Mason, S. T. Myers, A. C. S. Readhead, Astrophys. J.,
in press (see xxx.lanl.gov/abs/astro-ph/0101170). \\
13. G. P. Holder, J. J. Mohr, J. E. Carlstrom, A. E. Evrard, E.
M. Leitch, Astrophys. J. 544, 629 (2000). \\
14. Z. Haiman, J. J. Mohr, G. P. Holder, Astrophys. J., in press
(see xxx.lanl.gov/abs/astro-ph/0002336). \\
15. S. J. LaRoque et al., in preparation. \\
16. W. L. Holzapfel et al., Astrophys. J. 481, 35 (1997). \\
17. D. J. Fixsen, Astrophys. J. 473, 576 (1996).
\normalsize
\newpage
\begin{figure}[t]
\begin{center}
\hbox to \columnwidth{
\hskip0.2in
\hbox to 2.6in{\epsfxsize=2.6in\epsfbox{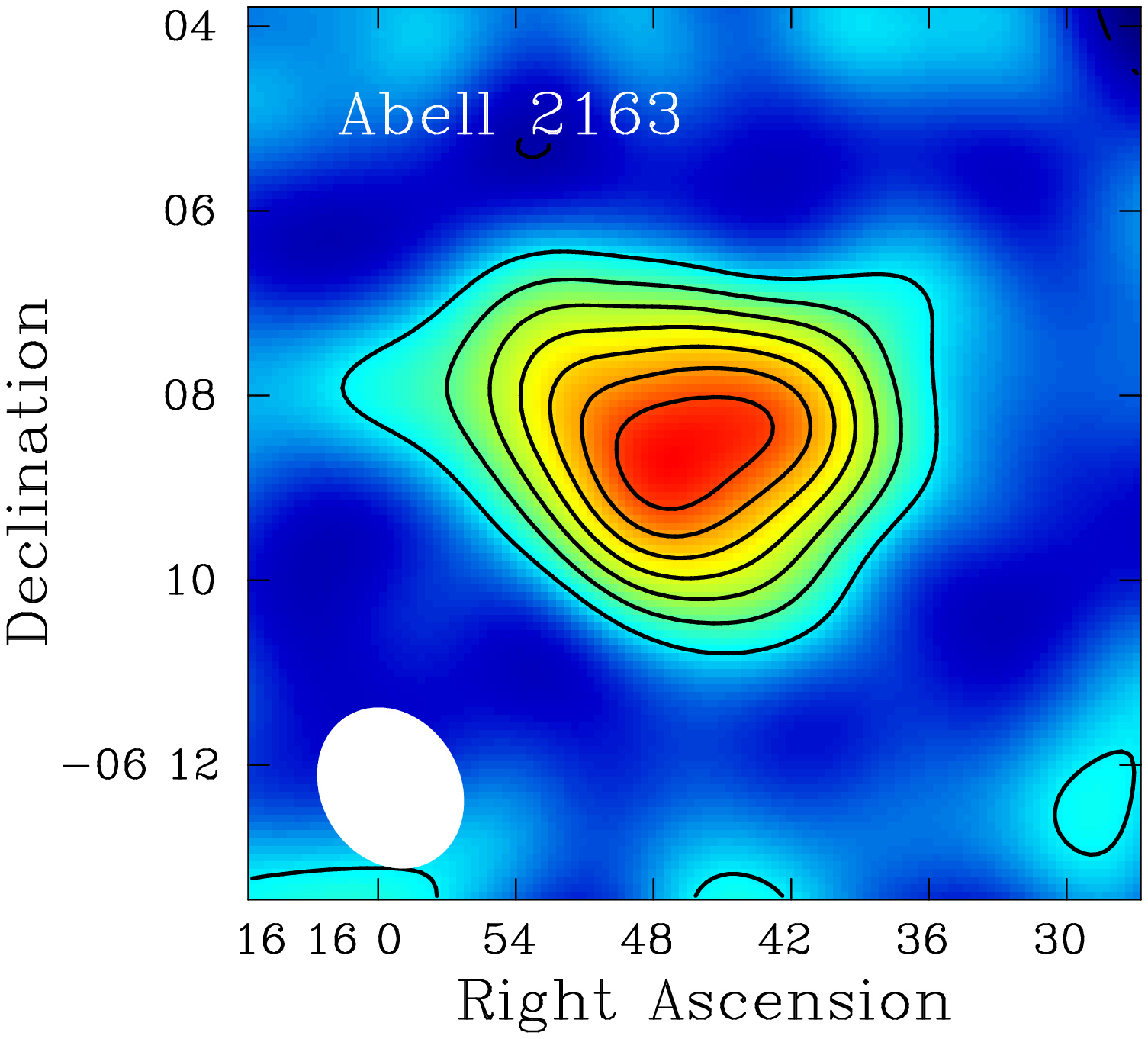}}
\hfil
\hbox to
3.1in{\epsfxsize=3.1in\epsfbox{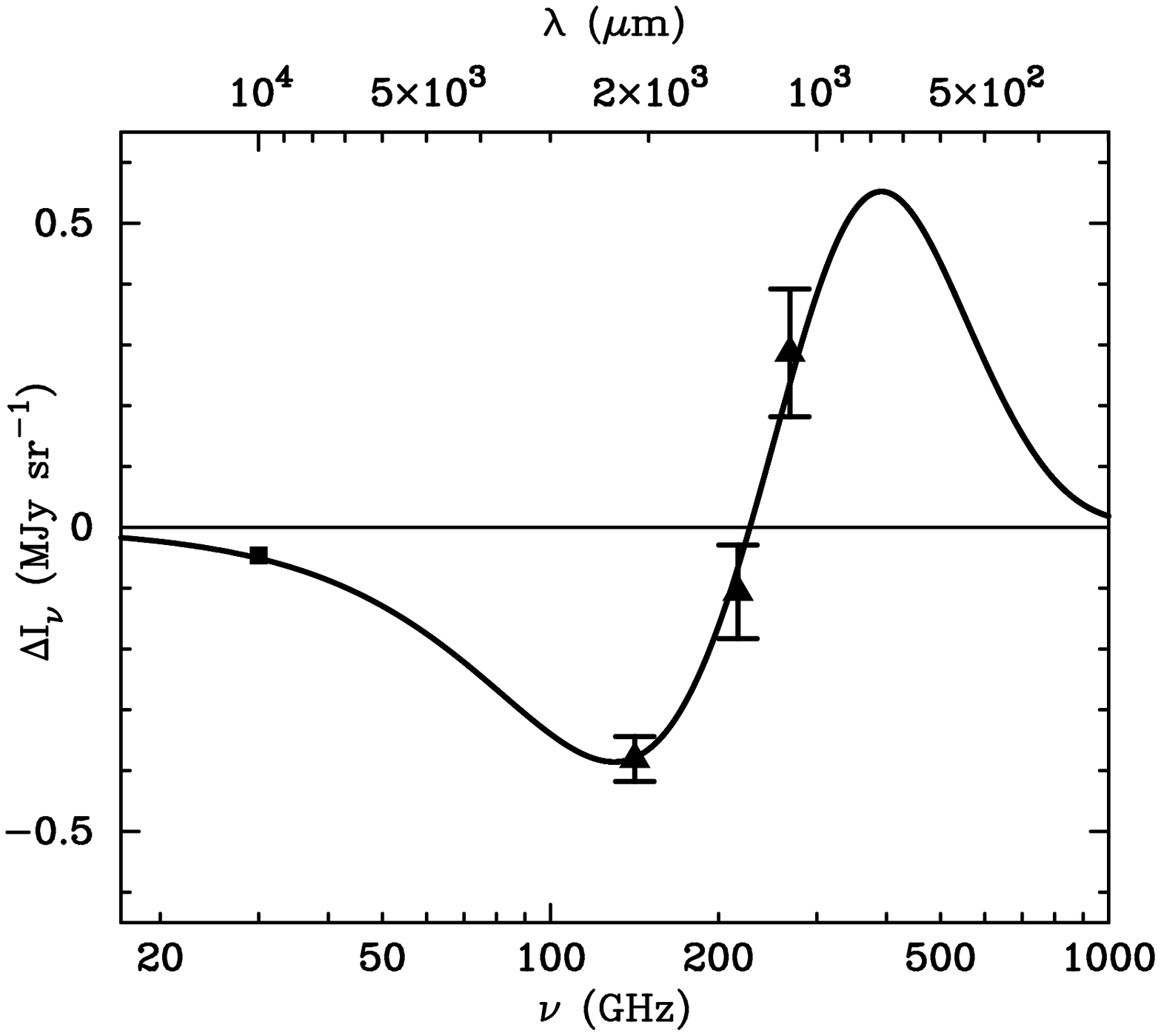}}
\hskip0.2in
}
\caption{
\small
The Sunyaev-Zel'dovich effect. (Left) An image of the SZ
effect toward the galaxy cluster Abell 2163 (obtained with the
Berkeley-Illinois-Maryland Association and the Caltech Owens
Valley interferometers) reveals a decrease in the otherwise
uniform brightness of the CMBR toward the galaxy cluster (15).
The contour interval is two times the root-mean-square noise in
the map. (Right) SZ effect spectral distortion (in units of
MJy/sr) relative to the undistorted CMBR spectrum measured
toward Abell 2163 (15, 16). On average, the CMBR photons
interacting with the hot cluster gas are shifted to higher
energy, resulting in a deficit of photons at low frequencies and
an excess at high frequencies relative to the CMBR spectrum
(17).
\normalsize}
\end{center}
\end{figure}

\vskip -30pt
\begin{figure}[b]
\epsscale{0.55}
\plotone{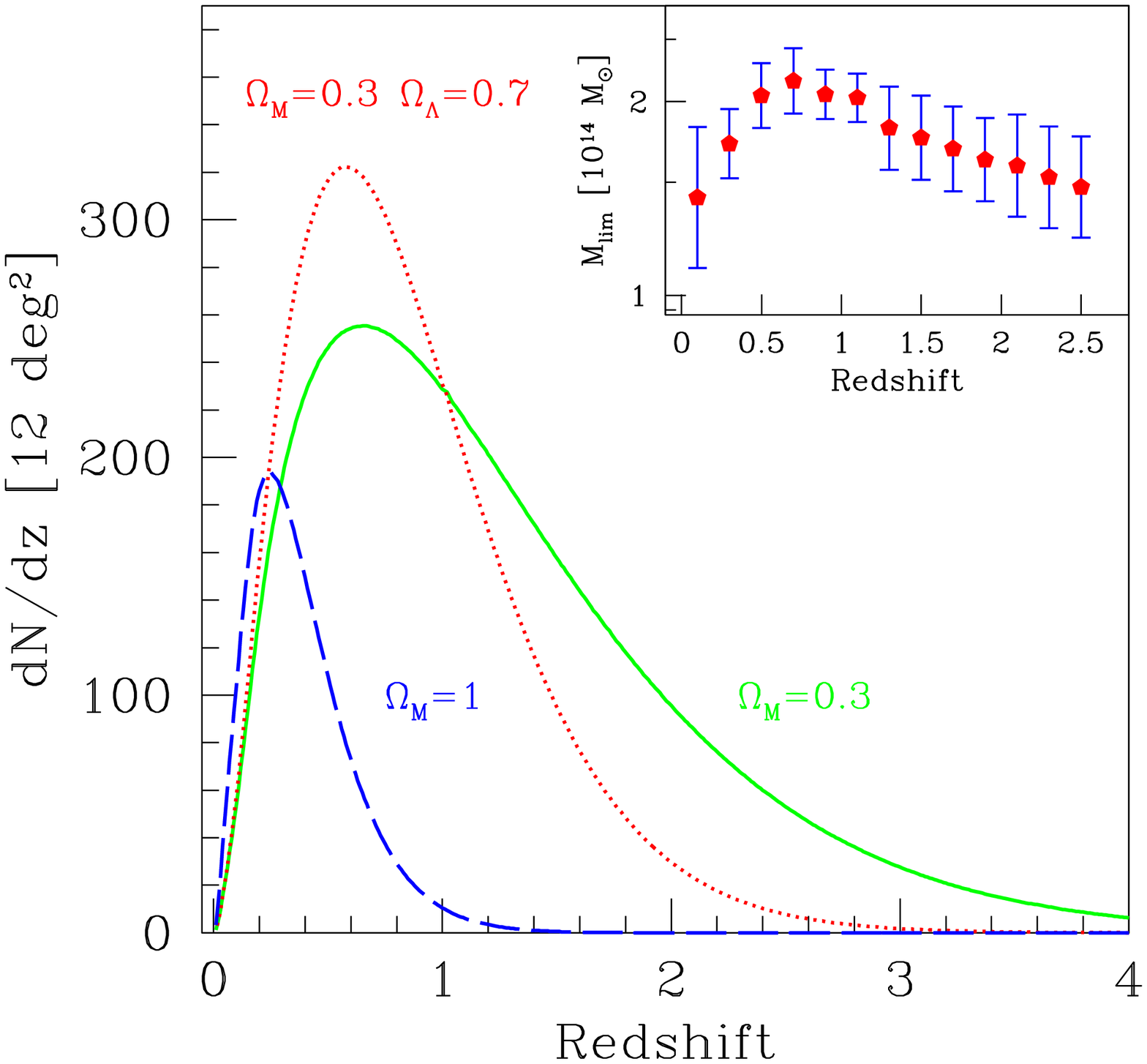}
\caption{
\small
Constraining cosmological models. The predicted
number density of clusters (dN/dz) detectable in a deep SZ effect
survey, calculated for various cosmological models (13). The mass
detection threshold of an SZ effect survey is insensitive to
redshift; all clusters of mass greater than $2.5 \times 10^{14}$
times the mass of the sun should be detectable, independent of
when they were formed.
\normalsize
}

\end{figure}

\end{document}